\documentclass[prd,a4,twocolumn,superscriptaddress,nofootinbib]{revtex4}
\usepackage{hyperref,amsmath,amssymb,epsfig,bm}
\usepackage{ifthen}

\begin{document}

\renewcommand{\eprint}[1]{\href{http://arxiv.org/abs/#1}{#1}}
\renewcommand{\bibinfo}[2]{\ifthenelse{\equal{#1}{isbn}}{%
\href{http://cosmologist.info/ISBN/#2}{#2}}{#2}}
\newcommand{\adsurl}[1]{\href{#1}{ADS}}

\newcommand{\Msun}{M_\odot}
\newcommand{\vrms}{v_{\text{rms}}}
\newcommand{\tot}{{\text{tot}}}
\newcommand{\ud}{{\text{d}}}
\newcommand{\Mpc}{\text{Mpc}}
\newcommand{\half}{{\textstyle \frac{1}{2}}}
\newcommand{\third}{{\textstyle \frac{1}{3}}}
\newcommand{\numfrac}[2]{{\textstyle \frac{#1}{#2}}}
\newcommand{\ra}{\rangle}
\newcommand{\la}{\langle}
\renewcommand{\d}{\text{d}}
\newcommand{\grad}{\nabla}
\newcommand{\km}{\rm{\,km\,}}
\newcommand{\rv}{r_{200}}

\newcommand{\begm}{\begin{pmatrix}}
\newcommand{\enm}{\end{pmatrix}}

\newcommand{\threej}[6]{{\begm #1 & #2 & #3 \\ #4 & #5 & #6 \enm}}
\newcommand{\fsky}{f_{\text{sky}}}
\newcommand{\arcmin}{\text{arcmin}}
\newcommand\Tr{{\rm Tr}}
\newcommand{\cla}{\mathcal{A}}
\newcommand{\clb}{\mathcal{B}}
\newcommand{\clc}{\mathcal{C}}
\newcommand{\cle}{\mathcal{E}}
\newcommand{\clf}{\mathcal{F}}
\newcommand{\clg}{\mathcal{G}}
\newcommand{\clh}{\mathcal{H}}
\newcommand{\cli}{\mathcal{I}}
\newcommand{\clj}{\mathcal{J}}
\newcommand{\clk}{\mathcal{K}}
\newcommand{\cll}{\mathcal{L}}
\newcommand{\clm}{\mathcal{M}}
\newcommand{\cln}{\mathcal{N}}
\newcommand{\clo}{\mathcal{O}}
\newcommand{\clp}{\mathcal{P}}
\newcommand{\clq}{\mathcal{Q}}
\newcommand{\clr}{\mathcal{R}}
\newcommand{\cls}{\mathcal{S}}
\newcommand{\clt}{\mathcal{T}}
\newcommand{\clu}{\mathcal{U}}
\newcommand{\clv}{\mathcal{V}}
\newcommand{\clw}{\mathcal{W}}
\newcommand{\clx}{\mathcal{X}}
\newcommand{\cly}{\mathcal{Y}}
\newcommand{\clz}{\mathcal{Z}}
\newcommand{\CMBFAST}{\textsc{cmbfast}}
\newcommand{\CAMB}{\textsc{camb}}
\newcommand{\Omtot}{\Omega_{\mathrm{tot}}}
\newcommand{\Omb}{\Omega_{\mathrm{b}}}
\newcommand{\Omc}{\Omega_{\mathrm{c}}}
\newcommand{\Omm}{\Omega_{\mathrm{m}}}
\newcommand{\omb}{\omega_{\mathrm{b}}}
\newcommand{\omc}{\omega_{\mathrm{c}}}
\newcommand{\omm}{\omega_{\mathrm{m}}}
\newcommand{\Omdm}{\Omega_{\mathrm{DM}}}
\newcommand{\Omnu}{\Omega_{\nu}}
\newcommand{\vpsi}{\mathbf{\psi}}
\renewcommand{\vr}{\mathbf{r}}

\newcommand{\vTheta}{\mathbf{\Theta}}
\newcommand{\vdelta}{\boldsymbol{\delta}}

\newcommand{\vU}{\mathbf{U}}
\newcommand{\vQ}{\mathbf{Q}}

\newcommand{\Oml}{\Omega_\Lambda}
\newcommand{\OmK}{\Omega_K}

\newcommand{\Hunit}{~\text{km}~\text{s}^{-1} \Mpc^{-1}}
\newcommand{\Gyr}{{\rm Gyr}}
\newcommand{\muK}{\mu\rm{K}}
\newcommand{\muKarcmin}{\,\muK\,\arcmin}

\newcommand{\nrun}{n_{\text{run}}}

\newcommand{\lmax}{l_{\text{max}}}

\newcommand{\zre}{z_{\text{re}}}
\newcommand{\mpl}{m_{\text{Pl}}}

\newcommand{\valpha}{{\boldsymbol{\alpha}}}
\newcommand{\vgrad}{{\boldsymbol{\nabla}}}

\newcommand{\vphi}{\mathbf{\psi}}
\newcommand{\vv}{\mathbf{v}}
\newcommand{\vd}{\mathbf{d}}
\newcommand{\vC}{\mathbf{C}}
\newcommand{\vT}{\mathbf{\Theta}}
\newcommand{\vX}{\mathbf{X}}
\newcommand{\vn}{\mathbf{n}}
\newcommand{\vy}{\mathbf{y}}
\newcommand{\mN}{\bm{N}}
\newcommand{\eV}{\,\text{eV}}
\newcommand{\vtheta}{\bm{\theta}}
\newcommand{\tT}{\tilde{T}}
\newcommand{\tE}{\tilde{E}}
\newcommand{\tB}{\tilde{B}}

\newcommand{\mCh}{\hat{\bm{C}}}
\newcommand{\Ch}{\hat{C}}

\newcommand{\Bt}{\tilde{B}}
\newcommand{\Et}{\tilde{E}}
\newcommand{\bld}[1]{\mathrm{#1}}
\newcommand{\mLambda}{\bm{\Lambda}}
\newcommand{\mA}{\bm{A}}
\newcommand{\mC}{\bm{C}}
\newcommand{\mQ}{\bm{Q}}
\newcommand{\mU}{\bm{U}}
\newcommand{\mX}{\bm{X}}
\newcommand{\mV}{\bm{V}}
\newcommand{\mP}{\bm{P}}
\newcommand{\mR}{\bm{R}}
\newcommand{\mW}{\bm{W}}
\newcommand{\mD}{\bm{D}}
\newcommand{\mI}{\bm{I}}
\newcommand{\mH}{\bm{H}}
\newcommand{\mM}{\bm{M}}
\newcommand{\mS}{\bm{S}}
\newcommand{\mzero}{\bm{0}}
\newcommand{\mL}{\bm{L}}

\newcommand{\btheta}{\bm{\theta}}
\newcommand{\bphi}{\bm{\psi}}

\newcommand{\vb}{\mathbf{b}}
\newcommand{\vA}{\mathbf{A}}
\newcommand{\vAt}{\tilde{\mathbf{A}}}
\newcommand{\ve}{\mathbf{e}}
\newcommand{\vE}{\mathbf{E}}
\newcommand{\vB}{\mathbf{B}}
\newcommand{\vEt}{\tilde{\mathbf{E}}}
\newcommand{\vBt}{\tilde{\mathbf{B}}}
\newcommand{\vEw}{\mathbf{E}_W}
\newcommand{\vBw}{\mathbf{B}_W}
\newcommand{\vx}{\mathbf{x}}
\newcommand{\vXt}{\tilde{\vX}}
\newcommand{\vXb}{\bar{\vX}}
\newcommand{\vTb}{\vTheta}
\newcommand{\vTt}{\tilde{\vT}}
\newcommand{\vY}{\mathbf{Y}}
\newcommand{\vBwr}{{\vBw^{(R)}}}
\newcommand{\RW}{{W^{(R)}}}

\newcommand{\mUt}{\tilde{\mU}}
\newcommand{\mVt}{\tilde{\mV}}
\newcommand{\mDt}{\tilde{\mD}}

\newcommand{\Rot}{\begm \mzero &\mI \\ -\mI & \mzero \enm}
\newcommand{\Pt}{\begm \vEt \\ \vBt \enm}

\newcommand{\edth}{\,\eth\,}
\renewcommand{\beth}{\,\overline{\eth}\,}

\newcommand{\sE}{{}_{|s|}E}
\newcommand{\sB}{{}_{|s|}B}
\newcommand{\sElm}{\sE_{lm}}
\newcommand{\sBlm}{\sB_{lm}}
\newcommand{\vnhat}{{\hat{\mathbf{n}}}}

\newcommand{\vk}{{\mathbf{k}}}
\newcommand{\vl}{{\mathbf{l}}}
\newcommand{\vlhat}{{\hat{\mathbf{l}}}}
\newcommand{\vL}{{\mathbf{L}}}
\newcommand{\vq}{{\mathbf{q}}}
\newcommand{\Cgl}{C_{\text{gl}}}
\newcommand{\Cgltwo}{C_{\text{gl},2}}

\title{Observational constraints and cosmological parameters}

\author{Antony Lewis}
\homepage{http://cosmologist.info}

 \affiliation{Institute of Astronomy, Madingley Road, Cambridge, CB3 0HA, UK.}

\begin{abstract}
I discuss the extraction of cosmological parameter constraints from the recent WMAP 3-year data, both on its own and in combination with other data. The large degeneracies in the first year data can be largely broken with the third year data, giving much better parameter constraints from WMAP alone. The polarization constraint on the optical depth is crucial to obtain the main results, including $n_s<1$ in basic six-parameter models. Almost identical constraints can also be obtained using only temperature data with a prior on the optical depth. I discuss the modelling of secondaries when extracting parameter constraints, and show that the effect of CMB lensing is about as important as SZ and slightly increases the inferred value of the spectral index. Constraints on correlated matter isocurvature modes are not radically better than before, and the data is consistent with a purely adiabatic spectrum. Combining WMAP 3-year data with data from the Lyman-$\alpha$ forest suggests somewhat higher values for $\sigma_8$ than from WMAP alone.
\end{abstract}

\maketitle

\section{Introduction}

The WMAP satellite has now provided three years worth of beautiful data giving a clear picture of the CMB temperature over most of the sky and allowing robust constraints on many cosmological parameters~\cite{Hinshaw:2006ia,Spergel:2006hy}. I shall briefly review parameter estimation methodology, then discuss in more detail topical issues relating to WMAP parameter constraints. I also discuss the utility of combining with other data, and show some example joint constraints obtained by combining with Lyman-$\alpha$ data. I shall assume familiarity with standard abbreviations and cosmological parameters.

\section{Parameter estimation}

The CMB sky is expected (and observed) to be closely Gaussian, and noise properties can also often be approximated as Gaussian. Given a set of parameters, linear-theory predictions for the CMB power spectrum can be computed quickly using numerical codes such as CAMB~\cite{Lewis:1999bs} and CMBFAST~\cite{Seljak:1996is}. Since the statistics of observed sky or power spectrum estimators are quite well understood, we can therefore compute the likelihood of a given set of parameters given the observed data. We then wish to calculate the posterior distribution of various parameters of interest, $P(\text{parameters}|\text{data})$, given a set of priors. In potentially large dimensional parameter spaces, the information in the posterior distribution is most easily extracted by using a set of samples from the distribution. For example an estimate of the posterior mean of a given parameter is given simply by the average of the value of that parameter in the set of samples. More complicated quantities can also be computed from samples, for example the correlations of different parameters, marginalized two-dimensional distributions, etc.

The methods almost universally adopted for sampling are variations of the Metropolis-Hastings Markov Chain Monte Carlo method. This method is explained in e.g. Refs.~\cite{Neal93,Christensen:2001gj,Lewis:2002ah}, and consists of constructing a way of exploring parameter space such that the value of the parameters at any given step are a sample from the posterior distribution. There is considerable choice in the details of how these methods are implemented, and some discussion of suitable step-proposal distributions is given in Ref.~\cite{cosmomc_notes,Dunkley:2004sv}. A fairly complete implementation aimed at sampling and analysing non-pathological distributions (i.e. preferably unimodal and not too distorted) most common in cosmology is publicly available in the CosmoMC\footnote{\url{http://cosmologist.info/cosmomc}} package~\cite{Lewis:2002ah} used here; see also AnalyseThis~\cite{Doran:2003ua}.

Once a set of samples is available, almost every statistical question of interest can be answered using estimators constructed from the samples. The probability density at a point in parameter space is given simply by the number density of samples. Marginalized probabilities in reduced dimensions are given simply by the number density of samples for the parameters in those reduced dimensions. Furthermore often new priors or data can be added very quickly using importance sampling: a way of re-weighting samples based on the difference of the new and old likelihoods --- see Ref.\cite{Lewis:2002ah} for further discussion. Chains for the WMAP 3-year analysis are available from the LAMBDA\footnote{\url{http://lambda.gsfc.nasa.gov/}} website, and are compatible with CosmoMC's \emph{GetDist} program.

\section{WMAP 3-year constraints}

It is well known from the WMAP first year data that with good measurements of only the first two acoustic peaks of the CMB temperature power spectrum there remains an important degeneracy between the spectral index $n_s$, the optical depth $\tau$, the baryon density $\Omega_b h^2$ and the amplitude of fluctuations parameterized by $A_s$ or $\sigma_8$. This degeneracy remains in the 3-year temperature data, as shown by the samples in Fig.~\ref{TT2D}.

\subsection{Constraining the optical depth}

\begin{figure}
\begin{center}
\psfig{figure=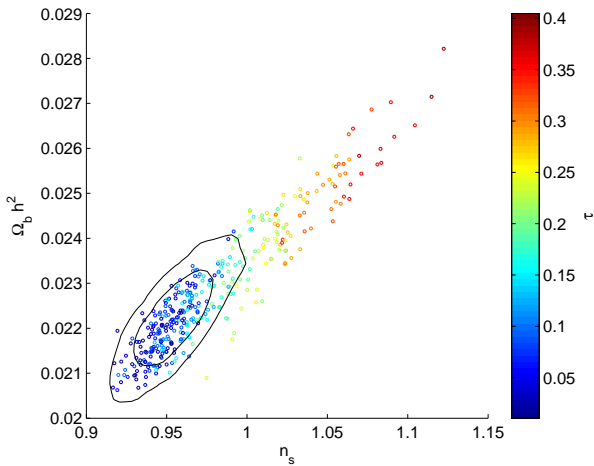,width=8cm}
\caption{Constraints from WMAP 3-year temperature (points) and joint with polarization ($68\%$ and $95\%$ contours) for a basic six parameter $\Lambda$CDM model (no tensors). The points represent samples from the posterior distribution, and are coloured by the value of the optical depth $\tau$. Polarization constrains the optical depth, breaking the main flat-model degeneracy and suggesting $n_s<1$.}
\label{TT2D}
\end{center}
\end{figure}

\begin{figure}
\begin{center}
\psfig{figure=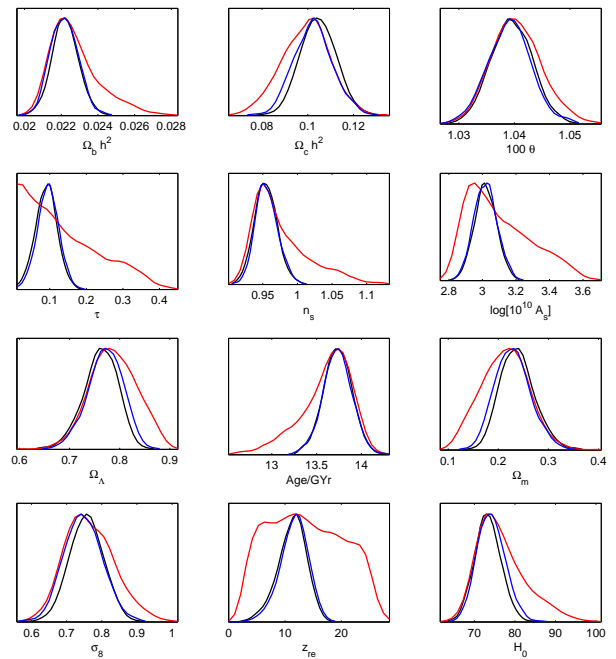,width=8cm}
\caption{Constraints from WMAP 3-year temperature (red), temperature and polarization (black), and temperature with a Gaussian prior on the optical depth $\tau=0.10\pm 0.03$ (blue). The top six parameters have flat priors and are sampled using MCMC, the bottom six parameters are derived.}
\label{TT1D}
\end{center}
\end{figure}
A large scale polarization signal on the CMB can be generated at reionization due to scattering of the CMB quadrupole. By measuring the large scale $E$-mode power spectrum, one can therefore constrain the optical depth. The 3-year WMAP analysis makes a heroic attempt to clean out dominating foregrounds to extract this signal~\cite{Page:2006hz}. Various consistency checks (for example $B$-mode power spectrum consistent with zero) suggest that this has been achieved to sufficient accuracy to reliably constrain the optical depth. However the possibility of some unidentified foreground remains, and it is unclear how large the foreground-induced systematic error in the result is (unlike for the temperature analysis, there is no marginalization over the foreground templates or fitting uncertainties).

Using the WMAP 3-year result for the large scale polarization the optical depth $\tau$ can be constrained, breaking the main flat-model degeneracy, and suggesting that $n_s<1$ in basic tensor-free models as shown in Fig.~\ref{TT2D}. In fact this is essentially the \emph{only} information in the polarization that effects basic cosmological model parameters. For example if we take a prior on the optical depth $\tau=0.10\pm 0.03$ (consistent with the polarization result) essentially identical parameter constraints are obtained using only the temperature results as shown in Fig.~\ref{TT1D}.  The polarization spectrum on smaller scales may however be useful for constraining extended models, for example with isocurvature modes. The improvement in parameter constraints from the better measurement of the temperature power spectrum with the 3-year data is relatively modest; using the same $\tau$ prior with the first year temperature power spectrum gives an almost identical constraint on $n_s$ as using the 3-year data, though the matter densities are constrained rather better due to the better measurement of the second and third acoustic peaks.

Different priors on the optical depth or reionization redshift can be applied very quickly to existing chains using importance sampling. For example, my personal prior is approximately a Gaussian with $\zre=8\pm 3$ truncated to zero at $\zre <6$, which is towards the lower end of the distribution obtained by WMAP (e.g. if there were an unidentified foreground remaining in the WMAP data, the value of the optical depth would be expected to come down further). In this case using WMAP temperature and polarization without tensors the $n_s<1$ result is very significant ($n_s < 0.98$ at 2-sigma) and the value of $\sigma_8$ comes down a bit to $0.73\pm 0.05$.

\subsection{Effect of secondaries}

\begin{figure}
\begin{center}
\psfig{figure=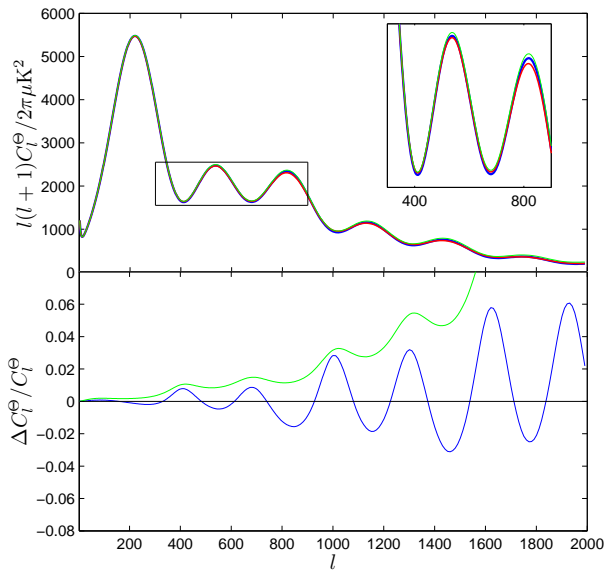,width=8cm}
\caption{The theoretical unlensed (black) and smoother lensed (red) CMB temperature power spectra (top) and the difference (bottom blue) for a fiducial WMAP 3-year $\Lambda$CDM model with $n_s=0.95$, $\tau=0.09$. Green lines are unlensed but include the SZ spectrum from Ref.~\cite{Komatsu:2002wc} evaluated for fiducial parameters as used by the WMAP team (with marginalized amplitude) at 22.8GHz. At higher frequencies the SZ contribution is somewhat smaller.}
\label{lensedT}
\end{center}
\end{figure}

\begin{figure}
\begin{center}
\psfig{figure=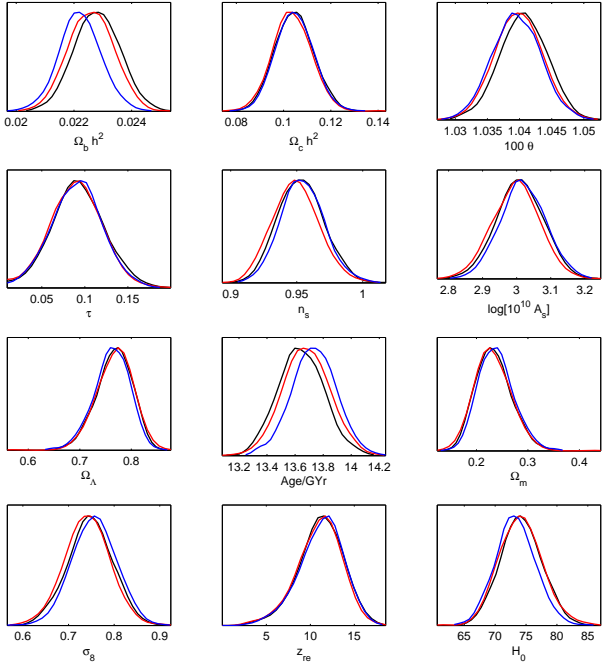,width=8cm}
\caption{Six parameter WMAP 3-yr constraints with no secondary modelling (blue), marginalizing over SZ model amplitude (red) and marginalizing over SZ and including CMB lensing (black).}
\label{lensSZ}
\end{center}
\end{figure}

The WMAP 3-year parameter analysis accounts for an SZ contribution by marginalizing over the amplitude of an analytic model calculated for a fixed fiducial set of parameters~\cite{Spergel:2006hy}. The effect of including SZ on the cosmological parameter constraints is not large, and comparable to that obtainable by changing priors in the base parameters. Since the SZ adds power on small scales, adding a contribution means that $n_s$ has to decrease slightly, shifting the posterior distribution further down from $n_s=1$. However when modelling secondaries, all relevant effects should be included. It is well known that CMB lensing also has a percent level effect on the relevant scales --- see Fig.~\ref{lensedT} (for a recent detailed review of CMB lensing see Ref.~\cite{Lewis:2006fu}). Lensing smooths out the acoustic peaks, and on WMAP scales the main effect is a reduction in the height of the third peak height, as well as smaller effects on the earlier peaks and troughs. The effect can be compensated by increasing $n_s$, so including lensing tends to \emph{increase} $n_s$. The effect is shown in Fig~\ref{lensSZ} and is comparable magnitude to SZ, but has an opposite effect on $n_s$. The best fit sample has a slightly better $\chi^2$ with lensing included, with $\Delta\chi^2\sim 1$.  Since the effect of secondaries is small, they do not have a large effect on the conclusions of the WMAP 3-year analysis. However if SZ is included it would seem sensible to also include CMB lensing. The CMB lensing effect can be computed very accurately and is trivial to include, and the effect on WMAP parameters shown here is compatible with the forecast given in Ref.~\cite{Lewis:2005tp}. The fact that SZ and lensing have nearly opposite effects suggests that in practice rather accurate WMAP constraints on $n_s$ can be obtained by neglecting both. However the effect on $\Omega_b h^2$ is in the same direction, and including both shifts the mean up by nearly one sigma compared to including neither, giving $\Omega_b h^2 = 0.0228\pm 0.0008$.

\subsection{Effect of likelihood estimation}

Though in principle the likelihood from a Gaussian CMB sky should be easy to calculate, in practice it can be rather difficult due to partial sky coverage, dataset size, non-uniform noise, beam uncertainty, etc.~\cite{Hinshaw:2006ia}. The 3-year WMAP likelihood approximation is much improved from the first-year one, including full covariance information on large scales where the $C_l$ are most non-Gaussian and noise correlations are most important.
Unless otherwise stated results quoted in this paper are computed using the likelihood code supplied by the WMAP team with default settings and no secondary modelling. However there are options in the likelihood code depending on how the effect of the beam uncertainty is modelled~\cite{Peiris:2006ug}. For basic six-parameter models, using the alternative options of Ref.~\cite{Peiris:2006ug} tends to shift $n_s$ to slightly higher values\footnote{\url{http://cosmocoffee.info/viewtopic.php?t=503}}, increasing $n_s$ by $\sim 0.006$. This suggests an estimate of the systematic error due to small scale likelihood modelling of $\clo(0.01)$ on $n_s$, comparable to the effect of secondaries and choice of variables with flat priors.

\section{Combining data}

\begin{figure}
\begin{center}
\psfig{figure=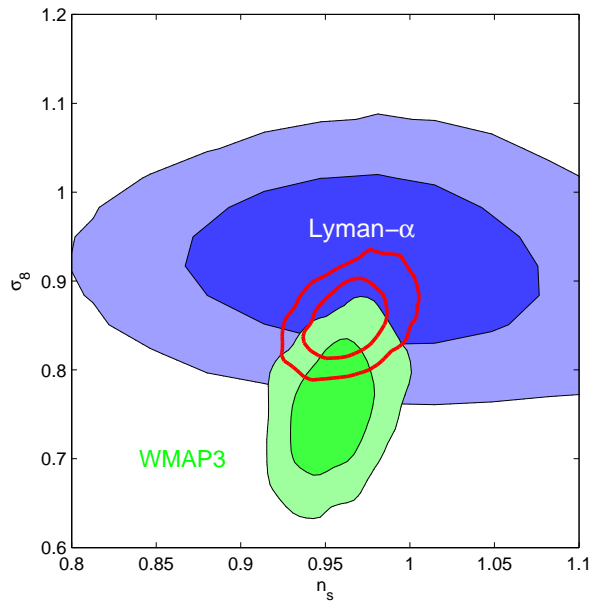,width=8cm}
\caption{Constraints on the spectral index $n_s$ and $\sigma_8$ from WMAP only (green), SDSS Lyman-$\alpha$ (blue, Ref.~\cite{McDonald:2004xn}), and the joint constraint (red). Contours enclose $68\%$ and $95\%$ of the probability, and the model is $\Lambda$CDM with no tensors. Lyman alpha only contours are for fixed best-fit joint values of the other parameters and dependent upon this prior: they  give a visual clue to the direction and amount by which the data pull the joint constraints, but the absolute position of the contours is fairly meaningless.}
\label{lya}
\end{center}
\end{figure}

\begin{figure}
\begin{center}
\psfig{figure=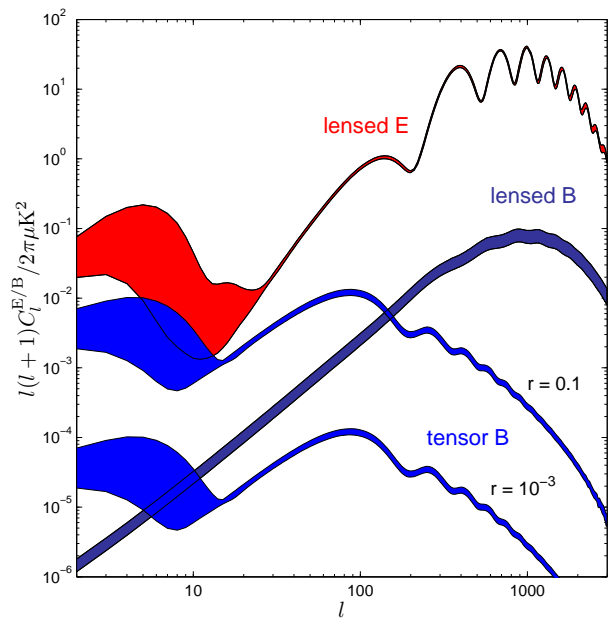,width=8cm}
\caption{The 95\%-confidence regions for the CMB polarization power spectra given WMAP 3-yr, Boomerang~\cite{Jones:2005yb} and 2dF~\cite{Percival:2001hw} data with an HST prior $h=0.72\pm 0.08$.
The parameters varied are those for a constant spectral index $\Lambda$CDM model with insignificant neutrino mass and sharp reionization. The two tensor results are the $B$-mode power spectra expected from for a scale-invariant tensor-mode power spectrum with $A_T= 4r \times 10^{-9}$ and two possible values of $r$. The effect of tensors on the cosmological parameter constraints was neglected, and a prior $\zre>6$ was assumed. This is an updated version of the figure in Ref.~\cite{Lewis:2006fu}.
\label{ClRange}}
\end{center}
\end{figure}

WMAP alone provides good constraints on many parameters, especially those that are directly related to the power spectrum. For example the ratio of the sound horizon at last scattering to the angular diameter distance to last scattering is constrained to $\sim 1\%$. Other combinations of parameter are also measured very well, for example WMAP alone gives the linear theory constraints
\begin{eqnarray}
\left( \frac{\Omega_m}{0.238}\right)\left(\frac{h}{0.73}\right)^{2.2} \left( \frac{e^{-\tau}\sigma_8}{0.689}\right)^{-0.9} &=& 1 \pm 0.01 \\
 \left( \frac{e^{-\tau}\sigma_8}{0.689}\right)\left(\frac{h}{0.73}\right)^{1.15}\left( \frac{\Omega_m}{0.238}\right)^{-0.066} &=& 1 \pm 0.04.
\end{eqnarray}
However the third combination of these parameters is only constrained to $20\%$. Any data that can improve on these constraints in any direction can be used to improve parameter constraints. In addition external data provides an important consistency check, and any deviations from consistency may indicate departures from the assumed cosmological model.

The WMAP results are now so good that for simple models the small scale $C_l$ can be predicted rather accurately from the WMAP data on larger scales. Data from other CMB experiments currently give relatively large errors on the smaller scale acoustic peaks, and the improvement to parameter constraints from including other CMB data on linear scales is now rather modest. To significantly improve parameter constraints from 3-year WMAP a tight constraint on the third acoustic peak would be very useful. As shown in Fig.~\ref{ClRange} small scale polarization observations have to be very sensitive to compete with the prediction that can be made from current data assuming a basic cosmological model, and it will be a while before small scale CMB $E$-polarization helps at all with constraining parameters in these models. However consistency with the predictions is an important consistency check.


The WMAP team give constraints with various combinations of data as discussed in Ref.~\cite{Spergel:2006hy}. Of particular note is the significant increase in the expected value of $\sigma_8$ when including data from weak lensing. Here as an example I consider combination with Lyman-$\alpha$ data from SDSS, Refs.~\cite{McDonald:2004xn}. This serves to measure the matter power spectrum scales much smaller than those directly probed by the CMB and hence has the potential to beak degeneracies involving $n_s$ and $\sigma_8$. The data from Ref.~\cite{Viel:2004bf} are less constraining that the SDSS results, so I concentrate on the latter here.

Figure~\ref{lya} shows the joint constrain from WMAP and Lyman-$\alpha$ without tensors. As with weak lensing, adding Lyman alpha increases the values of $\sigma_8$ from the value preferred by WMAP alone, with the SDSS joint constraint giving $\sigma_8 = 0.86 \pm 0.03$. The joint constraint is actually otherwise fairly consistent with WMAP alone, for example the best fit sample has $\tau = 0.09, n_s=0.97$, though it does favour higher values of $\Omega_m \sim 0.3$. It is consistent with $n_s<1$ at two sigma with no running of the spectral index. 

\section{Extended models}

\begin{figure}
\begin{center}
\psfig{figure=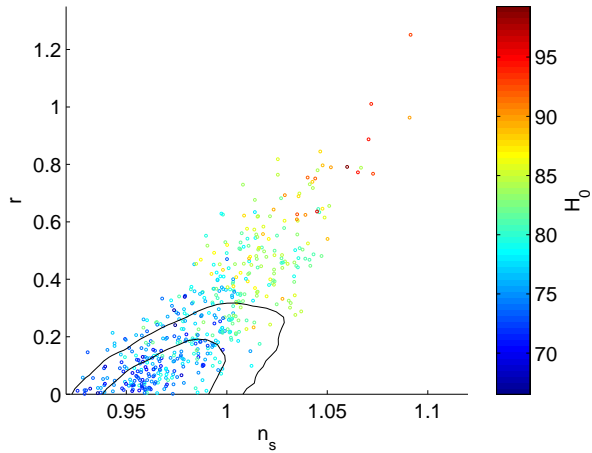,width=8cm}
\caption{Allowing tensors, samples from WMAP 3-year temperature and polarization (points), compared to the joint constraint with SDSS Lyman-$\alpha$ data from Ref.~\cite{McDonald:2004xn} ($68\%$ and $95\%$ contours).}
\label{tens}
\end{center}
\end{figure}

One of the most interesting extensions of the basic 6-parameter $\Lambda$CDM model is one allowing for a contribution from tensors modes as predicted by some models of inflation.
Figure~\ref{tens} shows constraints on the tensor-scalar ratio $r$ (defined in terms of the initial power spectra as in the WMAP papers) from WMAP alone\footnote{Note that the contours in the corresponding Fig. 14 of Ref.~\cite{Spergel:2006hy} were wrong in version 1 of that paper.} and when combined with Lyman-$\alpha$. Clearly there is no evidence for a tensor contribution, however allowing for tensors does allow the possibility of $n_s\ge 1$: if tensors add power to the low-$l$ multipoles, the spectral index can be bluer and still maintain the correct overall ratio of small and large scale power. Many future observations will be aimed at detecting the $B$-mode polarization signal from tensor modes, and Fig.~\ref{ClRange} shows the expected spectrum for a couple of possible amplitudes in comparison to the predicted signal from lensed scalar modes. Further discussions of inflationary constraints from WMAP 3-year data are given in Refs.~\cite{Spergel:2006hy,Peiris:2006ug,Alabidi:2006qa}.

\begin{figure}
\begin{center}
\psfig{figure=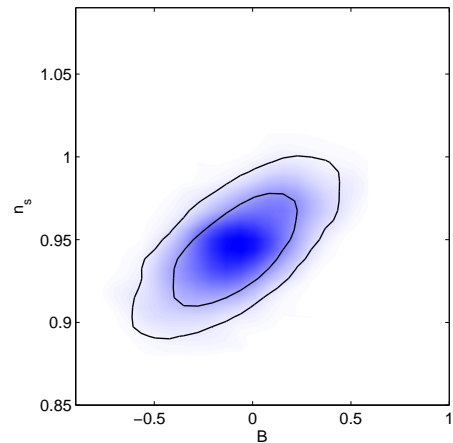,width=6cm}
\caption{Constraints on the correlated matter isocurvature mode ratio $B$ as defined in Ref.~\cite{Gordon:2002gv} using WMAP 3-year data combined with small scale CMB~\cite{Readhead:2004gy,Kuo:2002ua} and 2dF data~\cite{Percival:2001hw}.}
\label{curv}
\end{center}
\end{figure}

Another interesting possibility is an isocurvature mode contribution. Probably the simplest possibility is a totally correlated contribution from matter isocurvature modes, for example from a curvaton model of inflation. Here the spectral index is fixed to that of the adiabatic modes, and there is only one extra parameter, the ratio of isocurvature modes (which can be negative for anti-correlated modes). The updated constraint from Ref.~\cite{Gordon:2002gv} is shown in Fig.~\ref{curv}, corresponding to a $95\%$ confidence constraint $-0.42< B < 0.25$. As with older data there is no evidence at all for isocurvature modes, which should not be surprising given that a simple adiabatic model fits the data well. The WMAP 3-year data does not radically improve the constraint because the main effect is on large scales where the first year data was already cosmic variance limited in the temperature. However degeneracy breaking does help, as does improved sensitivity on the polarization. I expect that more general isocurvature analyses (e.g. updating Ref.~\cite{Bucher:2004an}) should have a similar conclusion: no evidence for isocurvature modes, though a significant fraction still allowed.

\section{Conclusions}

The WMAP 3-year results appear to show a remarkable agreement with a simple six parameter $\Lambda$CDM model. Numerous extended models have been discussed  by the WMAP team~\cite{Spergel:2006hy}, though none fit the data significantly better. There is some tension by between the $\sigma_8$ values favoured by WMAP and weak lensing and Lyman-alpha data, and this deserves further investigation. The WMAP data alone suggest that either $n_s<1$ or there is something else interesting,  for example a significant tensor mode component.

Future CMB observations that can provide an accurate measurement of the optical depth or third acoustic peak are needed to constrain parameters significantly better with CMB alone. Precision measurements of the temperature spectrum to small scales should allow the spectral index to be determined independently of the polarization constraint on the optical depth. Current and future small scale polarization observations are unlikely to improve basic parameter constraints significantly, though they can be a good test of alternative models. Detecting large scale $B$-polarization from gravitational waves is of course the main goal of many future missions, and would be a powerful way to constrain models of inflation.

CMB observations are in principle clean way to constrain early universe perturbations and many cosmological parameters. However they are rather insensitive to  other parameters, for example the late time evolution of the dark energy. To constrain these parameters complementary data are needed. Consistency with different data sets is a good check on the assumed cosmological model. If consistent other data sets can also be used to break partial degeneracies in the CMB constraints, giving better joint parameter estimates. Many additional data sets not considered here are analysed in combination with WMAP 3-year data in Ref.\cite{Spergel:2006hy}.

\section*{Acknowledgements}
I thank the organisers for inviting me to give this talk and an excellent conference. I thank  Matteo Viel, George Efstathiou, Hiranya Peiris and fellow conferees for useful discussion. I thank K. Abazajian and J. Lesgourgues for making their Lyman-$\alpha$ CosmoMC modules public.
I acknowledge the use of the Legacy Archive for Microwave Background Data Analysis\footnote{\url{http://lambda.gsfc.nasa.gov/}} (LAMBDA). Support for LAMBDA is provided by the NASA Office of Space Science. I thank CITA for use of their Beowulf computer. I am supported by a PPARC Advanced Fellowship.


\begin{thebibliography}{24}
\expandafter\ifx\csname natexlab\endcsname\relax\def\natexlab#1{#1}\fi
\expandafter\ifx\csname bibnamefont\endcsname\relax
  \def\bibnamefont#1{#1}\fi
\expandafter\ifx\csname bibfnamefont\endcsname\relax
  \def\bibfnamefont#1{#1}\fi
\expandafter\ifx\csname citenamefont\endcsname\relax
  \def\citenamefont#1{#1}\fi
\expandafter\ifx\csname url\endcsname\relax
  \def\url#1{\texttt{#1}}\fi
\expandafter\ifx\csname urlprefix\endcsname\relax\def\urlprefix{URL }\fi
\providecommand{\bibinfo}[2]{#2}
\providecommand{\eprint}[2][]{\url{#2}}

\bibitem[{\citenamefont{Hinshaw et~al.}(2006)}]{Hinshaw:2006ia}
\bibinfo{author}{\bibfnamefont{G.}~\bibnamefont{Hinshaw}} \bibnamefont{et~al.}
  (\bibinfo{year}{2006}), \eprint{astro-ph/0603451}.

\bibitem[{\citenamefont{Spergel et~al.}(2006)}]{Spergel:2006hy}
\bibinfo{author}{\bibfnamefont{D.~N.} \bibnamefont{Spergel}}
  \bibnamefont{et~al.} (\bibinfo{year}{2006}), \eprint{astro-ph/0603449}.

\bibitem[{\citenamefont{Lewis et~al.}(2000)\citenamefont{Lewis, Challinor, and
  Lasenby}}]{Lewis:1999bs}
\bibinfo{author}{\bibfnamefont{A.}~\bibnamefont{Lewis}},
  \bibinfo{author}{\bibfnamefont{A.}~\bibnamefont{Challinor}},
  \bibnamefont{and} \bibinfo{author}{\bibfnamefont{A.}~\bibnamefont{Lasenby}},
  \bibinfo{journal}{Astrophys. J.} \textbf{\bibinfo{volume}{538}},
  \bibinfo{pages}{473} (\bibinfo{year}{2000}), \eprint{astro-ph/9911177}.

\bibitem[{\citenamefont{Seljak and Zaldarriaga}(1996)}]{Seljak:1996is}
\bibinfo{author}{\bibfnamefont{U.}~\bibnamefont{Seljak}} \bibnamefont{and}
  \bibinfo{author}{\bibfnamefont{M.}~\bibnamefont{Zaldarriaga}},
  \bibinfo{journal}{Astrophys. J.} \textbf{\bibinfo{volume}{469}},
  \bibinfo{pages}{437} (\bibinfo{year}{1996}), \eprint{astro-ph/9603033}.

\bibitem[{\citenamefont{Neal}(1993)}]{Neal93}
\bibinfo{author}{\bibfnamefont{R.~M.} \bibnamefont{Neal}}
  (\bibinfo{year}{1993}), \bibinfo{note}{\url{http://cosmologist.info/Neal93}}.

\bibitem[{\citenamefont{Christensen et~al.}(2001)\citenamefont{Christensen,
  Meyer, Knox, and Luey}}]{Christensen:2001gj}
\bibinfo{author}{\bibfnamefont{N.}~\bibnamefont{Christensen}},
  \bibinfo{author}{\bibfnamefont{R.}~\bibnamefont{Meyer}},
  \bibinfo{author}{\bibfnamefont{L.}~\bibnamefont{Knox}}, \bibnamefont{and}
  \bibinfo{author}{\bibfnamefont{B.}~\bibnamefont{Luey}},
  \bibinfo{journal}{Class. Quant. Grav.} \textbf{\bibinfo{volume}{18}},
  \bibinfo{pages}{2677} (\bibinfo{year}{2001}), \eprint{astro-ph/0103134}.

\bibitem[{\citenamefont{Lewis and Bridle}(2002)}]{Lewis:2002ah}
\bibinfo{author}{\bibfnamefont{A.}~\bibnamefont{Lewis}} \bibnamefont{and}
  \bibinfo{author}{\bibfnamefont{S.}~\bibnamefont{Bridle}},
  \bibinfo{journal}{Phys. Rev.} \textbf{\bibinfo{volume}{D66}},
  \bibinfo{pages}{103511} (\bibinfo{year}{2002}), \eprint{astro-ph/0205436}.

\bibitem[{\citenamefont{Lewis and Bridle}()}]{cosmomc_notes}
\bibinfo{author}{\bibfnamefont{A.}~\bibnamefont{Lewis}} \bibnamefont{and}
  \bibinfo{author}{\bibfnamefont{S.}~\bibnamefont{Bridle}},
  \bibinfo{note}{\url{http://cosmologist.info/notes/cosmomc.ps.gz}}.

\bibitem[{\citenamefont{Dunkley et~al.}(2005)\citenamefont{Dunkley, Bucher,
  Ferreira, Moodley, and Skordis}}]{Dunkley:2004sv}
\bibinfo{author}{\bibfnamefont{J.}~\bibnamefont{Dunkley}},
  \bibinfo{author}{\bibfnamefont{M.}~\bibnamefont{Bucher}},
  \bibinfo{author}{\bibfnamefont{P.~G.} \bibnamefont{Ferreira}},
  \bibinfo{author}{\bibfnamefont{K.}~\bibnamefont{Moodley}}, \bibnamefont{and}
  \bibinfo{author}{\bibfnamefont{C.}~\bibnamefont{Skordis}},
  \bibinfo{journal}{Mon. Not. Roy. Astron. Soc.}
  \textbf{\bibinfo{volume}{356}}, \bibinfo{pages}{925} (\bibinfo{year}{2005}),
  \eprint{astro-ph/0405462}.

\bibitem[{\citenamefont{Doran and Mueller}(2004)}]{Doran:2003ua}
\bibinfo{author}{\bibfnamefont{M.}~\bibnamefont{Doran}} \bibnamefont{and}
  \bibinfo{author}{\bibfnamefont{C.~M.} \bibnamefont{Mueller}},
  \bibinfo{journal}{JCAP} \textbf{\bibinfo{volume}{0409}}, \bibinfo{pages}{003}
  (\bibinfo{year}{2004}), \eprint{astro-ph/0311311}.

\bibitem[{\citenamefont{Page et~al.}(2006)}]{Page:2006hz}
\bibinfo{author}{\bibfnamefont{L.}~\bibnamefont{Page}} \bibnamefont{et~al.}
  (\bibinfo{year}{2006}), \eprint{astro-ph/0603450}.

\bibitem[{\citenamefont{Komatsu and Seljak}(2002)}]{Komatsu:2002wc}
\bibinfo{author}{\bibfnamefont{E.}~\bibnamefont{Komatsu}} \bibnamefont{and}
  \bibinfo{author}{\bibfnamefont{U.}~\bibnamefont{Seljak}},
  \bibinfo{journal}{Mon. Not. Roy. Astron. Soc.}
  \textbf{\bibinfo{volume}{336}}, \bibinfo{pages}{1256} (\bibinfo{year}{2002}),
  \eprint{astro-ph/0205468}.

\bibitem[{\citenamefont{Lewis and Challinor}(2006)}]{Lewis:2006fu}
\bibinfo{author}{\bibfnamefont{A.}~\bibnamefont{Lewis}} \bibnamefont{and}
  \bibinfo{author}{\bibfnamefont{A.}~\bibnamefont{Challinor}}
  (\bibinfo{year}{2006}), \eprint{astro-ph/0601594}.

\bibitem[{\citenamefont{Lewis}(2005)}]{Lewis:2005tp}
\bibinfo{author}{\bibfnamefont{A.}~\bibnamefont{Lewis}},
  \bibinfo{journal}{Phys. Rev.} \textbf{\bibinfo{volume}{D71}},
  \bibinfo{pages}{083008} (\bibinfo{year}{2005}), \eprint{astro-ph/0502469}.

\bibitem[{\citenamefont{Peiris and Easther}(2006)}]{Peiris:2006ug}
\bibinfo{author}{\bibfnamefont{H.}~\bibnamefont{Peiris}} \bibnamefont{and}
  \bibinfo{author}{\bibfnamefont{R.}~\bibnamefont{Easther}}
  (\bibinfo{year}{2006}), \eprint{astro-ph/0603587}.

\bibitem[{\citenamefont{Viel et~al.}(2004)\citenamefont{Viel, Haehnelt, and
  Springel}}]{Viel:2004bf}
\bibinfo{author}{\bibfnamefont{M.}~\bibnamefont{Viel}},
  \bibinfo{author}{\bibfnamefont{M.~G.} \bibnamefont{Haehnelt}},
  \bibnamefont{and} \bibinfo{author}{\bibfnamefont{V.}~\bibnamefont{Springel}},
  \bibinfo{journal}{Mon. Not. Roy. Astron. Soc.}
  \textbf{\bibinfo{volume}{354}}, \bibinfo{pages}{684} (\bibinfo{year}{2004}),
  \eprint{astro-ph/0404600}.

\bibitem[{\citenamefont{McDonald et~al.}(2005)}]{McDonald:2004xn}
\bibinfo{author}{\bibfnamefont{P.}~\bibnamefont{McDonald}}
  \bibnamefont{et~al.}, \bibinfo{journal}{Astrophys. J.}
  \textbf{\bibinfo{volume}{635}}, \bibinfo{pages}{761} (\bibinfo{year}{2005}),
  \eprint{astro-ph/0407377}.

\bibitem[{\citenamefont{Jones et~al.}(2005)}]{Jones:2005yb}
\bibinfo{author}{\bibfnamefont{W.~C.} \bibnamefont{Jones}} \bibnamefont{et~al.}
  (\bibinfo{year}{2005}), \eprint{astro-ph/0507494}.

\bibitem[{\citenamefont{Percival et~al.}(2001)}]{Percival:2001hw}
\bibinfo{author}{\bibfnamefont{W.~J.} \bibnamefont{Percival}}
  \bibnamefont{et~al.}, \bibinfo{journal}{MNRAS}
  \textbf{\bibinfo{volume}{327}}, \bibinfo{pages}{1297} (\bibinfo{year}{2001}),
  \eprint{astro-ph/0105252}.

\bibitem[{\citenamefont{Alabidi and Lyth}(2006)}]{Alabidi:2006qa}
\bibinfo{author}{\bibfnamefont{L.}~\bibnamefont{Alabidi}} \bibnamefont{and}
  \bibinfo{author}{\bibfnamefont{D.~H.} \bibnamefont{Lyth}}
  (\bibinfo{year}{2006}), \eprint{astro-ph/0603539}.

\bibitem[{\citenamefont{Gordon and Lewis}(2003)}]{Gordon:2002gv}
\bibinfo{author}{\bibfnamefont{C.}~\bibnamefont{Gordon}} \bibnamefont{and}
  \bibinfo{author}{\bibfnamefont{A.}~\bibnamefont{Lewis}},
  \bibinfo{journal}{Phys. Rev.} \textbf{\bibinfo{volume}{D67}},
  \bibinfo{pages}{123513} (\bibinfo{year}{2003}), \eprint{astro-ph/0212248}.

\bibitem[{\citenamefont{Readhead et~al.}(2004)}]{Readhead:2004gy}
\bibinfo{author}{\bibfnamefont{A.~C.~S.} \bibnamefont{Readhead}}
  \bibnamefont{et~al.}, \bibinfo{journal}{Astrophys. J.}
  \textbf{\bibinfo{volume}{609}}, \bibinfo{pages}{498} (\bibinfo{year}{2004}),
  \eprint{astro-ph/0402359}.

\bibitem[{\citenamefont{Kuo et~al.}(2004)}]{Kuo:2002ua}
\bibinfo{author}{\bibfnamefont{C.-l.} \bibnamefont{Kuo}} \bibnamefont{et~al.}
  (\bibinfo{collaboration}{ACBAR}), \bibinfo{journal}{Astrophys. J.}
  \textbf{\bibinfo{volume}{600}}, \bibinfo{pages}{32} (\bibinfo{year}{2004}),
  \eprint{astro-ph/0212289}.

\bibitem[{\citenamefont{Bucher et~al.}(2004)\citenamefont{Bucher, Dunkley,
  Ferreira, Moodley, and Skordis}}]{Bucher:2004an}
\bibinfo{author}{\bibfnamefont{M.}~\bibnamefont{Bucher}},
  \bibinfo{author}{\bibfnamefont{J.}~\bibnamefont{Dunkley}},
  \bibinfo{author}{\bibfnamefont{P.~G.} \bibnamefont{Ferreira}},
  \bibinfo{author}{\bibfnamefont{K.}~\bibnamefont{Moodley}}, \bibnamefont{and}
  \bibinfo{author}{\bibfnamefont{C.}~\bibnamefont{Skordis}},
  \bibinfo{journal}{Phys. Rev. Lett.} \textbf{\bibinfo{volume}{93}},
  \bibinfo{pages}{081301} (\bibinfo{year}{2004}), \eprint{astro-ph/0401417}.

\end{thebibliography}
\providecommand{\aj}{Astrophys. J. }\providecommand{\apj}{Astrophys. J.
  }\providecommand{\apjl}{Astrophys. J. }\providecommand{\mnras}{MNRAS}

\end{document}